\documentclass[journal=jacsat,manuscript=article]{achemso}

\usepackage{chemformula} 
\usepackage[T1]{fontenc} 
\pdfoutput=1 

\newcommand*\mycommand[1]{\texttt{\emph{#1}}}

\author{Andrey Polyakov}
\author{Katayoon Mohseni}
\affiliation[MPI Halle]
{Max-Planck-Institut f\"{u}r Mikrostrukturphysik, Weinberg 2, D-06120 Halle (Germany)}

\author{Roberto Felici}
\affiliation[SPIN]
{Consiglio Nazionale delle Ricerche - SPIN, Via del Politecnico, 1
Roma 00133, (Italy)}

\author{Christian Tusche}
\affiliation[FZJ]
{Forschungszentrum J\"ulich, Peter Gr\"unberg Institut (PGI-6) 52425 J\"ulich (Germany)}
\alsoaffiliation [UNIDUE]
{Fakult\"at für Physik, Universit\"at Duisburg-Essen, 47057 Duisburg (Germany)}

\author{Ying-Jiun ~Chen}
\affiliation[FZJ]
{Forschungszentrum J\"ulich, Peter Gr\"unberg Institut (PGI-6) 52425 J\"ulich (Germany)}
\alsoaffiliation [UNIDUE]
{Fakult\"at für Physik, Universit\"at Duisburg-Essen, 47057 Duisburg (Germany)}

\author{Vitaliy Feyer}
\affiliation[FZJ]
{Forschungszentrum J\"ulich, Peter Gr\"unberg Institut (PGI-6) 52425 J\"ulich (Germany)}
\alsoaffiliation [UNIDUE]
{Fakult\"at für Physik, Universit\"at Duisburg-Essen, 47057 Duisburg (Germany)}

\author{Jochen Geck}
\affiliation[TUDD]
{Insitut für Festk\"{o}rper- und Materialphysik, Technische Universit\"{a}t Dresden, 01062 Dresden (Germany)}
\alsoaffiliation [CTQMAT]
{W\"urzburg-Dresden Cluster of Excellence ct.qmat, Technische Universit\"{a}t Dresden, 01062 Dresden (Germany)}

\author{Tobias Ritschel}
\affiliation{Insitut für Festk\"{o}rper und Materialphysik, Technische Universit\"{a}t Dresden, Haeckelstraße 3, D-01069 Dresden (Germany)}

\author{Juan Rubio-Zuazo}
\affiliation[ESRF]
{SpLine, Spanish CRG BM25 Beamline at the ESRF (The European Synchrotron), F-38000 Grenoble (France)}

\author{German R. Castro}
\affiliation[ESRF]
{SpLine, Spanish CRG BM25 Beamline at the ESRF (The European Synchrotron), F-38000 Grenoble (France)}

\author{Holger L. Meyerheim}
\affiliation[MPI Halle]
{Max-Planck-Institut f\"{u}r Mikrostrukturphysik, Weinberg 2, D-06120 Halle (Germany)}
\email{holger.meyerheim@mpi-halle.mpg.de}
\phone{+49 (0)345 5582633}
\fax{+49 (0)345 5511223}

\author{Stuart S. P. Parkin}
\affiliation[MPI Halle]
{Max-Planck-Institut f\"{u}r Mikrostrukturphysik, Weinberg 2, D-06120 Halle (Germany)}

\title[An \textsf{achemso} demo]
{Fermi surface chirality induced in a TaSe$_{2}$ monosheet formed by a Ta/Bi$_{2}$Se$_{3}$ interface reaction}

\abbreviations{IR,NMR,UV}
\keywords{American Chemical Society, \LaTeX}
\date{30. September 2020}
\begin{document}







\newpage

\begin{abstract}

Spin-momentum locking in topological insulators and materials with Rashba-type interactions is an extremely attractive feature for novel spintronic devices and is therefore under intense investigation. Significant efforts are underway to identify new material systems with spin-momentum locking, but also to create heterostructures with new spintronic functionalities. In the present study we address both subjects and investigate a van der Waals-type heterostructure consisting of the topological insulator Bi$_{2}$Se$_{3}$ and a single Se-Ta-Se triple-layer (TL) of H-type TaSe$_{2}$ grown by a novel method which exploits an interface reaction between the adsorbed metal and selenium. We then show, using surface x-ray diffraction, that the symmetry of the TaSe$_{2}$-like TL is reduced from D$_{3h}$ to C$_{3v}$ resulting from a vertical atomic shift of the tantalum atom. Spin- and angle-resolved photoemission indicates that, owing to the symmetry lowering, the states at the Fermi surface acquire an in-plane spin component forming a surface contour with a helical Rashba-like spin texture, which is coupled to the Dirac cone of the substrate. Our approach provides a new route to realize novel chiral two-dimensional electron systems via interface engineering that do not exist in the corresponding bulk materials.
\end{abstract}
\newpage
\section{Introduction}

Novel functionalities in potential spintronic devices especially involve the locking of the electron's spin and momentum as realized in Topological Insulators (TIs), as well as in Dirac and Weyl semimetals. The chiral topologic surface state (TSS) in TIs has been found to be very effective in converting a charge current into a spin current which can exert large spin-orbit torques (SOT) in an adjacent ferromagnetic (FM) layer~\cite{Mellnik2014,WangY2015,Fan2014,Kondou2016}. A critical issue is that the SOT efficiency resulting from the TSS can be influenced by several factors, such as the presence of bulk states, and the band-bending induced appearance of a two-dimensional electron gas, which recently has been shown to be minimized by reducing the Bi$_{2}$Se$_{3}$ film thickness ~\cite{Wang2017}. Similarly, transition metal dichalcogenides (TMDCs) with a non-trivial electronic structure containing a heavy metal such as Mo, W, Pt and Pd also have found remarkable interest as spin-source materials from significant charge to spin conversion ~\cite{Yuan2013,MacNeill2016,Clark2018,Xu2020}.

By contrast, the metallic TMDC TaSe$_{2}$ has a trivial electronic structure in its bulk form and crystallizes in the 2H structure (trigonal-prismatic coordination around tantalum by selenium). In this study we demonstrate that in a van der Waals (vdW) type heterostructure consisting of a \emph{single} Se-Ta-Se triple-layer (TL) on the (0001) surface of the TI Bi$_{2}$Se$_{3}$ a chirality is created in the Fermi surface electronic states. The TaSe$_{2}$ monosheet is prepared using a novel and simple method. This method, which does not rely on exfoliation or molecular beam epitaxy methods, that have been used in many previous studies, but rather uses an interface reaction between tantalum atoms directly deposited onto a Bi$_{2}$Se$_{3}$(0001) substrate.  We find that this simple method leads to flat islands that are formed from monosheets of well-ordered TaSe$_{2}$ with an H-type structure. The absence of the inversion center in the monosheet, in combination with the strong spin-orbit coupling (SOC) that is inherent to TaSe$_{2}$, results in a spin-splitting of the electronic states at the Fermi energy with oppositely spin-polarized states at the non-symmetry related K and K' points in the Brilluouin zone (BZ). The SOC lifts the spin-degeneracy of the bands inducing a spin-polarization which pins the electron spins to the \emph{out of plane} direction. This scenario is referred to as an "Ising-SOC". Until now, it has been generally assumed that monosheets of TMDC's are bulk-like. Surface xray diffraction (SXRD) analysis provides clear evidence for the vertical relaxation of the central tantalum atom in the prismatic selenium environment thereby lifting the horizontal mirror plane and lowering the point group symmetry from D$_{3h}$ to the C$_{3v}$. We then used spin- and momentum resolved photoemission spectroscopy, in combination with ab-initio calculations, to study the effect of the structural relaxation on the electronic structure.  We find a very important consequence is that the spin-polarized states at the Fermi surface (FS) acquire an in-plane spin component, in this way creating a chirality. Such a low-symmetry monosheet may serve as an efficient spin-source material not only avoiding difficulties encountered by bulk and free electron states in Bi$_{2}$Se$_{3}$ but also enabling a more sophisticated out of plane SOT to manipulate perpendicularly magnetized ferromagnetic films, as recently demonstrated in a WTe$_{2}$/permalloy heterostructure~\cite{MacNeill2016}

\newpage

\section{Results and discussion}

\subsection{Structure Analysis}

The TaSe$_{2}$ monosheet was prepared by depositing a sub-monolayer amount of tantalum on a pristine surface of a Bi$_{2}$Se$_{3}$(0001) single crystal, followed by annealing at 480$^{\circ}$C for several minutes.
Fig.~\ref{LEED} (a) and (b) show low energy electron diffraction (LEED) patterns of the pristine and TaSe$_{2}$ covered Bi$_{2}$Se$_{3}$(0001). The pristine Bi$_{2}$Se$_{3}$(0001) surface exhibits a threefold symmetric LEED pattern indicating the absence of crystal twinning. After preparation of the TaSe$_{2}$ monosheet, the substrate spots are attenuated and new spots appear that are related to the formation of a TaSe$_{2}$ monosheet [see Fig.~\ref{LEED} (b)]. The in-plane lattice parameter is derived from the spot position at approximately 1.19 reciprocal lattice units (r.l.u.) relative to the first order Bi$_{2}$Se$_{3}$ substrate spots. We find a$_{0}$=b$_{0}$=348~pm, which corresponds to a tensile strain of 1.5\% as compared to that of bulk 2H-TaSe$_{2}$ (343~pm)~\cite{Brown:1965}.

The tantalum covered surface was studied by scanning tunneling microscopy (STM) as outlined in Figure~\ref{STM} (a) and (b).  The Bi$_{2}$Se$_{3}$(0001) surface is characterized by terraces several hundred nanometers wide, which are separated by steps, 950 pm in height, that corresponds well with the thickness of a single quintuple layer (QL) of  Bi$_{2}$Se$_{3}$. The structure of Bi$_{2}$Se$_{3}$ is composed of QLs, each consisting of Se-Bi-Se-Bi-Se layers, that are separated by a van-der-Waals (vdW) gap. The QLs themselves are stacked in an A-B-C-A...~sequence. In the constant current STM image in Fig~\ref{STM}(a) (U$ =~-1$~V, I~$=100$~pA) TaSe$_{2}$ islands on the terraces appear as bright elevations with an apparent height of approximately 600~pm. The profile along the white line is given in Figure~\ref{STM}(b). It reflects the 950~pm high QL steps and the 600~pm high islands. Based on the STM images, the islands can be attributed to a monosheet of TaSe$_{2}$, whose height is expected to be approximately equal to 600~pm, i.e. half the unit cell lattice parameter c$_{0}$=1.271~nm of bulk 2H-TaSe$_{2}$~\cite{Brown:1965}. Based on these observations it can be concluded that the interface between the TaSe$_{2}$ triple layer and the first Bi$_{2}$Se$_{3}$ QL is characterized by the Se-Ta-Se/Se-Bi-Se-Bi-Se layer sequence, i.e. the interface is vdW-like. This conclusion is supported by the lattice parameters of the film being close to that of bulk TaSe$_{2}$ with an incommensurate film to substrate relationship and by the observation that sample annealing beyond 480$^{\circ}$C leads to the evaporation of the film. Based on the STM image alone, no unambiguous assignment of the film structure to H- or T-type can be made, since the height of the polyhedron is nearly identical in both polytypes~\cite{Yan2015}.

Detailed structural characterization was carried out by SXRD at the beamline BM25b of the European Synchrotron Radiation Facility (ESRF) in Grenoble (France) using a six-circle UHV diffractometer. In Fig.~\ref{Rods} the experimental intensities, $\mid$I(hk$\ell$)$\mid$$^{2}$, along several rods in reciprocal space are plotted on a log-scale versus momentum transfer (q$_{z}$) normal to the sample surface. In total, 86 symmetry independent reflection intensities were collected. In contrast to bulk crystals, the coordinate $\ell$=q$_{z}$/c$^{\star}$ in reciprocal lattice is a continuous (non-integer) parameter owing to the missing lattice periodicity along the c-axis of the monosheet. The reciprocal lattice unit refers to the the Bi$_{2}$Se$_{3}$ substrate where c$^{\star}$=1/c$_{0}$=1/(2.864~nm)=0.349~nm$^{-1}$.
A continuously varying intensity distribution is observed, reflecting the presence of an ultra-thin structure along the surface normal. The wide "bell-shaped" intensity profile, which is observed along all rods can be viewed as a finite size broadened Bragg-reflection from the monosheet.

Quantitative analysis was carried out by fitting the experimental intensities to the observed ones. Owing to the high symmetry of the crystal structure, that belongs to the plane group $p3m1$, the structure analysis is straightforward. All atoms are located at high symmetry positions as follows: Se at (0,0,0) [Wyckoff position (1a)], Ta at (1/3,2/3, z) [Wyckoff position (1b)]~\cite{Tables} and the top-layer Se atom at (0,0,z). The only free positional parameters are the z-positions of the tantalum and the second selenium atom at the top of the Se-Ta-Se triple layer. In addition, an overall scale factor and a Debye-parameter (B=8$\pi$<u>$^{2}$), reflecting thermal and static disorder~\cite{Kuhs1992}, were allowed to vary.

Solid lines in Fig.~\ref{Rods} represent the calculated intensities based on the structure model sketched in Fig:~\ref{Structure}(a). The fit quality is measured by the Goodness of Fit (GOF) parameter and the un-weighted residuum (Ru) ~\cite{RuI}. We derive values of GOF=1.56 and Ru=0.16. These values can be seen as very satisfactory. We note that the simulation also takes into account the presence of two rotational domains of the TaSe$_{2}$ unit cell with respect to the trigonal substrate surface. This is done by calculating the incoherent average of the structure factor magnitudes related to each of the two mutually 60$^{\circ}$ oriented domains. The twinning of the film structure is also the reason why the corresponding LEED spots exhibit a sixfold rather than a threefold symmetry. Fig.~\ref{Structure}(a) shows a schematic diagram of the structure derived from the optimal fit.

Small (red) and large (grey) balls represent selenium and tantalum atoms, respectively. The H-type monosheet is only one TL thick and stacked as in the bulk. Numbers indicate distances in picometers (pm) with those in brackets referring to bulk values in 2H-TaSe$_{2}$~\cite{Brown:1965}. The most important result is that the central tantalum atom is not located in the center of the prism at z=0.5 in relative lattice units (l.u.) but is rather shifted vertically to z=0.42 l.u. Simultaneously, the height of the TL is slightly increased from 336~pm~in the bulk to 337~pm in the monosheet [see numbers in brackets in Fig.~\ref{Structure}(a)]. The downward shift of the tantalum atom by about 27~pm out of the center position involves modifications of the interatomic distances also. The nearest Ta-Se distance, which in bulk 2H-TaSe$_{2}$ is equal to 259~pm, is modified to 280 and 246~pm corresponding to the upper and lower selenium atoms, respectively. We have carefully evaluated the uncertainty of the distance determination as shown in Fig.~\ref{Structure}(b) where the GOF is plotted versus the height (H) of the TL and the relative position (z) of the tantalum atom. The cross marks the global minimum at H=337 pm and z=0.42. Each contour level represents a step of 3\% with respect to the previous level. The uncertainty of H and z can be estimated by the variation of the GOF upon variation of H and z. An increase of the GOF by 3\% over the minimum is viewed as significant, especially here since the basic structure is unaffected by the very small modifications of H and z and because of the large number of independent reflections (86) relative to the number of free parameters (only two positional and one ADP).
Under this assumption we find uncertainties of $\Delta H= \pm$ 5~pm and $\Delta$z=$\pm$0.02 lattice units, the latter corresponding to $\pm$7 pm. As a consequence of the atomic shift, the symmetry of the H-type prism is lowered from  D$_{3h}$ ($\overline{6}$2m) to C$_{3v}$ (3m), which allows for the appearance of an in-plane component of the spin-polarization, as will be shown by spin- and momentum resolved photoemission experiments and by ab-initio calculations in the following.\\

\subsection{Electronic Structure}

Spin- and momentum resolved photoemission experiments  using a spin-resolving momentum microscope \cite{tus15} were carried out at the NanoESCA beamline~\cite{SCHNEIDER2012330} of the Elettra Synchrotron in Trieste (Italy). The sample was kept at 80~K and p-polarized light and a photon energy of $h\nu$=40~eV were used. The incident photon beam lies in the $k_{y}-k_{z}$ plane at an angle of 25$^\circ$ above the surface plane. By utilizing spin-resolving momentum microscopy, a wide acceptance angle of photoelectrons up to $\pm$90$^\circ$ can be simultaneously collected. The measured momentum discs provide a comprehensive access to the spin-resolved electronic states in two-dimensional ($k_x$, $k_y$) reciprocal-space momentum maps of the photoemission intensity throughout the entire surface Brillouin zone (SBZ). Furthermore, by scanning the binding energy $E_{B}$, the complete information of three-dimensional $E_{B}$($k_x$, $k_y$) maps, containing band dispersion along all directions of the SBZ, can be obtained \cite{SUGA2015119}.

{Figure~\ref{ARPES1} (a) and (b) show the spin-averaged band structure of the pristine and the TaSe$_{2}$ covered Bi$_{2}$Se$_{3}$(0001) surface along directions between the high-symmetry points of the first SBZ. The energy range lies between E$_{F}$ and 2.0 eV binding energy (E$_{B}$). The pristine surface exhibits a well-resolved Topological Surface State (TSS), namely a "Dirac cone" with the Dirac point (D$_{P}$) located about 350 meV below E$_{F}$, that is related to n-doping due to selenium defects. The TSS is stable upon formation of the TaSe$_{2}$ monosheet [Figure~\ref{ARPES1}(b)], but the D$_{P}$ is shifted downwards in energy by approximately 150 meV, indicating further n-doping, possibly by a higher concentration of selenium defects or by charge transfer from the film to the substrate.
The observed band structure of  TaSe$_{2}$ monosheet is similar to that of bulk 2H-TaSe$_{2}$, verifying that the Se-Ta-Se TL is indeed of the H-type. However, there are some differences, for instance, the number of bands crossing E$_{F}$ as discussed in several previous studies~\cite{Yan2015,Ryu2018,Ge2012}. The TaSe$_{2}$ monosheet is metallic, there is a Ta-5d related band which crosses E$_{F}$ between $\overline{\Gamma}$ and $\overline{M}$. Deeper lying and also strongly dispersive bands in the binding energy range between 1 and 2~eV are related to Se-p states~\cite{Ge2012}. The spin-resolved band structure of the TaSe$_{2}$ covered sample shown in Figure~\ref{ARPES1}(c) reveals that there is an antiparallel alignment between the spin texture of the TSS and that of the Ta-5d derived states which cross E$_{F}$ between the $\overline{M}$ and the $\overline{\Gamma}$ point.

Figures~\ref{ARPESCALC}(a-c) compare the experimental spin-integrated momentum map at the Fermi Level (E$_{F}$) with the calculated FS within the first surface Brillouin Zone (SBZ). The FS of the TaSe$_{2}$ monosheet is characterized by circular hole pockets around the $\overline{\Gamma}$- and the $\overline{K}$-point as well as the "dog-bone" like electron pocket around the $\overline{M}$-point, the latter being a consequence of spin-orbit coupling (SOC) and the lack of inversion symmetry in the TaSe$_{2}$ monosheet ~\cite{Yan2015,Ge2012}. Calculated momentum maps are superimposed on the experimental ones for z=0.38, z=0.43 and z=0.50. Direct inspection clearly reveals that the details of the FS sensitively depend on the z-position of the tantalum atom. While for z=0.38, the $\overline{K}$-pocket is too close to the $\overline{\Gamma}$ point, it moves outward until the vertical tantalum position reaches z=0.43. At this point the experimental and the calculated FS fit fairly well, both with regard to size and location of the pockets. Increasing z beyond 0.43 only leads to an increase of the size of the $\overline{K}$ hole pocket, rather than to a further outward shift. This is demonstrated in more detail in Figures~\ref{ARPESCALC}(d-h) showing the shift and the increase in size with increasing z. The optimum match between calculation and experiment is found for z=0.43, which is in perfect agreement with the SXRD analysis.

In non-magnetic solids the generation of spin-polarized electronic states requires the breaking of the global symmetry. The transition from bulk 2H-TaSe$_{2}$ with its inversion symmetric P6$_{3}$/mmc space group (D$_{6h}$ point group) to the "unrelaxed" 1H-TaSe$_{2}$ monosheet with $p3m1$ plane group symmetry (absence of inversion symmetry) represents such a case. The SOC lifts the spin-degeneracy of the bands inducing a spin-polarization which pins the electron spins to the \emph{out of plane} direction. This scenario is referred to as "Ising-SOC" and is regarded as being responsible for the strong enhancement of the critical field H$_{C2}$ in superconducting TMDC's such as in monosheet thick H-NbSe$_{2}$~\cite{Bawden2016,Zhou2016,He2018,Sohn2018} and gated MoS$_{2}$~\cite{Lu2015,Saito2015}. Most importantly, in the H-TaSe$_{2}$ monosheet studied here, the point group symmetry is further lowered from D$_{3h}$, to C$_{3v}$ by the vertical shift of the tantalum atom.

In consequence of the reduced symmetry of the TaSe$_{2}$ monosheet a pronounced in-plane spin-polarization is observed. The spin-resolved momentum map at the FS is displayed in Fig.~\ref{ARPESIPL}(a). It shows the in-plane projected spin-polarization which is coded as parallel (red) and antiparallel (blue) to the y direction, being the same as the direction of the incident beam.  The spin-resolved momentum-map shows that the TaSe$_{2}$ monosheet related states at E$_{F}$ exhibit a pronounced in-plane spin polarization, which is a direct consequence of its low-symmetry structure. In order to investigate this scenario in more detail we have calculated the spin-resolved states at E$_{F}$ using the fully relativistic Dirac-Kohn-Sham formalism as implemented in the FPLO18 package~\cite{Koepernik1999,Eschrig2004}. We carried out these calculations for different z-positions of the tantalum atom within the prism. Figure~\ref{ARPES1}(b) and (c) shows the results for z=0.43 and 0.50, respectively. While the latter refers to the "ideal" D$_{3h}$ symmetric prismatic polyhedron, the first one corresponds to the SXRD derived structure model. The spin-polarization of the states are displayed as projected along the y-axis allowing a direct comparison with experiment. In agreement with the the previous discussion there is no in-plane component of the spin texture for z=0.50. The situation changes if the tantalum atom is allowed to relax to 0.43 (Figure~\ref{ARPES1}(b)]. An in-plane component of the spin texture appears giving rise to a chirality which is opposite to that of the Dirac cone. Comparison between experiment and calculations shows a remarkable correspondence. For instance, in the experimental FS there is a chirality in the circular hole pocket around the $\overline{\Gamma}$-point as well as in the "dog-bone" like electron pocket around the $\overline{M}$-points. Also the circular state around the K-points is observable, albeit with a weak spectral weight.

Exploiting a simple interface reaction between tantalum adsorbed on the (0001) surface of Bi$_{2}$Se$_{3}$ we have prepared a heterostructure interfacing the TSS of Bi$_2$Se$_3$ with a two-dimensional H-type TaSe$_2$-monosheet. The combination of the Dirac-type states of the Bi$_{2}$Se$_{3}$ substrate with the SOC split electronic states at E$_{F}$ of the symmetry reduced H-TaSe$_2$ monosheet in its proximity causes a spin-momentum locking in the latter.
Our approach provides a new route to realize novel chiral two-dimensional electron systems that do not exist in the corresponding bulk materials and may open new approaches in the field of spin to charge conversion and spin-orbit torques ~\cite{Soumyanarayanan2016}. A non-magnetic TMDC in the ultra-thin film limit appears as advantageous in achieving a large SOT as it avoids the difficulties encountered with Bi$_{2}$Se$_{3}$ caused by the appearance of bulk and free electron states directly in contact with the ferromagnetic layer. Also, interesting transport phenomena may arise from the opposite chirality of the states at the $\overline{\Gamma}$- point of the adjacent films. This certainly needs to be explored in future studies. Notwithstanding these exciting possibilities for future research on this particular system, it is important to point out that the method described here can be used to prepare various selenium-based TMDC-layers on Bi$_2$Se$_3$. Our approach can therefore be used to prepare and explore various new and exciting materials and heterostructures.

\section{Methods}

\subsection{Sample preparation}

The experiments were carried out in-situ under ultra-high-vacuum (UHV) conditions in different UHV chambers. The (0001) surface of the Bi$_{2}$Se$_{3}$ single crystals were cleaned by repeated cycles of mild Ar$^{+}$ ion sputtering followed by annealing up to 500$^{\circ}$C for several minutes until Auger electron spectroscopy (AES) did not show any traces of carbon and oxygen~\cite{Roy.prb2014,Cavallin:2016} and a clear low energy electron diffraction (LEED) pattern was observed. Tantalum was deposited by evaporation from a tantalum rod heated by electron beam bombardment. The amount of tantalum deposited was calibrated by AES, scanning tunneling microscopy (STM) and ex-posteriori by SXRD. TaSe$_{2}$ is formed by annealing the as prepared sample at 480$^{\circ}$C for several minutes until the LEED diffraction pattern shows extra spots related to the formation of the TaSe$_{2}$ phase.

\subsection{Surface X-Ray Diffraction}

The XRD experiments were carried out at the beamline BM25B of the European Synchrotron Radiation Facility (ESRF) using a six circle diffractometer operated in the z-axis mode. The angle of the primary beam ($\lambda$=0.71~\AA) was set to $\alpha_{i}$=2.0 degrees.
Integrated reflections intensities [INT$_{obs}$(hkl)] were collected by performing omega scans about the sample normal and collecting the reflected beam by using an two-dimensional pixel detector. The INT$_{obs}$(hkl) were then multiplied by instrumental correction factors C$_{corr}$ as outlined in detail in Refs.~\cite{Schamper1993,Vlieg1997} yielding the experimental intensities [I$_{obs}$ (hkl)] via: I$_{obs}$(hkl)=INT$_{obs}$(hkl)$\times$ C$_{corr}$. The total uncertainty (1$\sigma$) of the I$_{obs}$(hkl)  is estimated to about 10\% as derived from the reproducibility of symmetry equivalent reflections.

\subsection{Spin-and momentum resolved photoemission spectroscopy}

Spin- and momentum resolved photoemission experiments using a Spin-resolving momentum microscope \cite{tus15} were carried out at the NanoESCA beamline~\cite{SCHNEIDER2012330} of the Elettra Synchrotron in Trieste (Italy). The sample was kept at 80~K, and photoelectrons from the TaSe$_{2}$ valence band region were excited by using p-polarized light and a photon energy of $h\nu$=40~eV.  Photoelectrons emitted into the complete solid angle above the sample surface were collected using a momentum microscope~\cite{Tusche2015}. The momentum microscope directly forms an image of the distribution of photoelectrons as function of the lateral crystal momentum (k$_{x}$,k$_{y}$) that is recorded
on an imaging detector. Here, we refer to these 2D intensity maps as “momentum
discs”, representing constant energy cuts through the valence band spectral function I(k$_{x}$,k$_{y}$,E). For spin-resolved momentum microscopy measurements, a W(100) single crystal is introduced into the momentum microscope, such that electrons are specularly reflected at the crystal surface such that the 2D momentum image is preserved \cite{Tusche2011}.
Electrons reflected from the crystal surface undergo spin-dependent scattering, such that the spin-polarization can be obtained at every point in the momentum discs over the entire SBZ \cite{Tusche2013}.The recorded intensity on the spin-resolving image detector branch depends on the electron spin being aligned parallel or antiparallel to the vertical quantization axis.

\subsection{First principle simulations}%
All density functional theory calculations have been performed using the FPLO18
package~\cite{Koepernik1999}. We modeled the Se-Ta-Se monolayer as
freestanding using the experimental lattice parameters with 20\AA{}  of vacuum in between layers in order to diminish
interactions between periodic images. Brillouin zone integration was performed
on a grid of $12\times12\times1$ irreducible $k$-points using the tetrahedron method. We employed
the fully relativistic four-component Dirac-Kohn-Sham theory as implemented in
FPLO18~\cite{Eschrig2004}. The generalized gradient approximation  (GGA) as parametrized
by Perdew, Burke, and Ernzerhof (Ref. \cite{Perdew1996}) was used to approximate the
exchange-correlation potential. Spin textures represent the $k$-dependent
expectation value of the spin operator.

\begin{acknowledgement}

This work is supported by the German Science foundation under Grant SPP 1666 (Topological Insulators) and by the BMBF under Grants 5K19PGA and 05K16PGB. TR gratefully acknowledges financial support from the German Research Foundation under Grant RI 2908/1-1.
TR and JG have been supported by the Deutsche Forschungsgemeinschaft through the SFB 1143 and the Würzburg-Dresden Cluster of Excellence EXC 2147 (ct.qmat).
The authors thank F. Weiss for technical assistance. HLM, AP, KM and RF thank the staff of the ESRF for their support and hospitality during their stay in Grenoble.
The authors thank the staff of the ESRF and Elettra for their help and hospitality during their visit in Grenoble and Trieste, and G. Zamborlini and M. Jugovac (PGI-6) for assistance, in using the NanoESCA beamline.

\end{acknowledgement}

\section*{Author contributions statement}

SSPP and HLM devised the experiments. AP investigated the sample preparation and characterization by STM and LEED. HLM, KM, GC, RF, AP and J.R.-Z. carried out the X-ray diffraction experiments at the ESRF. Data analysis was done by KM and HLM. The momentum microscopy experiments at Elettra were carried out by CT, VF, HLM, AP and YJC. TR and JG performed the first principle simulations. HLM, SSPP and CT wrote the manuscript. All authors have read and approved the manuscript.

\newpage
\bibliography{TaSe2}

\providecommand{\latin}[1]{#1}
\makeatletter
\providecommand{\doi}
  {\begingroup\let\do\@makeother\dospecials
  \catcode`\{=1 \catcode`\}=2 \doi@aux}
\providecommand{\doi@aux}[1]{\endgroup\texttt{#1}}
\makeatother
\providecommand*\mcitethebibliography{\thebibliography}
\csname @ifundefined\endcsname{endmcitethebibliography}
  {\let\endmcitethebibliography\endthebibliography}{}
\begin{mcitethebibliography}{37}
\providecommand*\natexlab[1]{#1}
\providecommand*\mciteSetBstSublistMode[1]{}
\providecommand*\mciteSetBstMaxWidthForm[2]{}
\providecommand*\mciteBstWouldAddEndPuncttrue
  {\def\EndOfBibitem{\unskip.}}
\providecommand*\mciteBstWouldAddEndPunctfalse
  {\let\EndOfBibitem\relax}
\providecommand*\mciteSetBstMidEndSepPunct[3]{}
\providecommand*\mciteSetBstSublistLabelBeginEnd[3]{}
\providecommand*\EndOfBibitem{}
\mciteSetBstSublistMode{f}
\mciteSetBstMaxWidthForm{subitem}{(\alph{mcitesubitemcount})}
\mciteSetBstSublistLabelBeginEnd
  {\mcitemaxwidthsubitemform\space}
  {\relax}
  {\relax}

\bibitem[Mellnik \latin{et~al.}(2014)Mellnik, Lee, Richardella, Grab, Mintun,
  Fischer, Vaezi, Manchon, Kim, Samarth, and Ralph]{Mellnik2014}
Mellnik,~A.~R.; Lee,~J.~S.; Richardella,~A.; Grab,~J.~L.; Mintun,~P.~J.;
  Fischer,~M.~H.; Vaezi,~A.; Manchon,~A.; Kim,~E.-A.; Samarth,~N.; Ralph,~D.~C.
  Spin-transfer torque generated by a topological insulator. \emph{Nature}
  \textbf{2014}, \emph{511}, 449\relax
\mciteBstWouldAddEndPuncttrue
\mciteSetBstMidEndSepPunct{\mcitedefaultmidpunct}
{\mcitedefaultendpunct}{\mcitedefaultseppunct}\relax
\EndOfBibitem
\bibitem[Wang \latin{et~al.}(2015)Wang, Deorani, Banerjee, Koirala, Brahlek,
  Oh, and Yang]{WangY2015}
Wang,~Y.; Deorani,~P.; Banerjee,~K.; Koirala,~N.; Brahlek,~M.; Oh,~S.; Yang,~H.
  Topological Surface States Originated Spin-Orbit Torques in
  ${\mathrm{Bi}}_{2}{\mathrm{Se}}_{3}$. \emph{Phys. Rev. Lett.} \textbf{2015},
  \emph{114}, 257202\relax
\mciteBstWouldAddEndPuncttrue
\mciteSetBstMidEndSepPunct{\mcitedefaultmidpunct}
{\mcitedefaultendpunct}{\mcitedefaultseppunct}\relax
\EndOfBibitem
\bibitem[Fan \latin{et~al.}(2014)Fan, Upadhyaya, Kou, Lang, Takei, Wang, He,
  Montazeri, Yu, Jiang, Nie, Schwartz, Tserkovnyak, and Wang]{Fan2014}
Fan,~Y.; Upadhyaya,~P.; Kou,~X.; Lang,~M.; Takei,~S.; Wang,~J.,~Zhenxing~Tang;
  He,~L.-T.,~Liang~Chang; Montazeri,~M.; Yu,~G.; Jiang,~W.; Nie,~T.;
  Schwartz,~R.~N.; Tserkovnyak,~Y.; Wang,~K.~L. Magnetization switching through
  giant spin-orbit torque in a magnetically doped topological insulator
  heterostructure. \emph{Nature Materials} \textbf{2014}, \emph{13}, 699\relax
\mciteBstWouldAddEndPuncttrue
\mciteSetBstMidEndSepPunct{\mcitedefaultmidpunct}
{\mcitedefaultendpunct}{\mcitedefaultseppunct}\relax
\EndOfBibitem
\bibitem[Kondou \latin{et~al.}(2016)Kondou, Yoshimi, Tsukazaki, Fukuma,
  Matsuno, Takahashi, Kawasaki, and Otani]{Kondou2016}
Kondou,~K.; Yoshimi,~R.; Tsukazaki,~A.; Fukuma,~Y.; Matsuno,~J.;
  Takahashi,~K.~S.; Kawasaki,~Y.,~M.~Tokura; Otani,~Y. Fermi-level-dependent
  charge-to-spin current conversion by Dirac surface states of topological
  insulators. \emph{Nature Physics} \textbf{2016}, \emph{12}, 1027\relax
\mciteBstWouldAddEndPuncttrue
\mciteSetBstMidEndSepPunct{\mcitedefaultmidpunct}
{\mcitedefaultendpunct}{\mcitedefaultseppunct}\relax
\EndOfBibitem
\bibitem[Wang \latin{et~al.}(2017)Wang, Zhu, Wu, Yang, Yu, Ramaswamy, Mishra,
  Shi, Elyasi, Teo, Wu, and Yang]{Wang2017}
Wang,~Y.; Zhu,~D.; Wu,~Y.; Yang,~Y.; Yu,~J.; Ramaswamy,~R.; Mishra,~R.;
  Shi,~S.; Elyasi,~M.; Teo,~K.-L.; Wu,~Y.; Yang,~H. Room temperature
  magnetization switching in topological insulator-ferromagnet heterostructures
  by spin-orbit torques. \emph{Nature Communication} \textbf{2017}, \emph{8},
  1364\relax
\mciteBstWouldAddEndPuncttrue
\mciteSetBstMidEndSepPunct{\mcitedefaultmidpunct}
{\mcitedefaultendpunct}{\mcitedefaultseppunct}\relax
\EndOfBibitem
\bibitem[Yuan \latin{et~al.}(2013)Yuan, Bahramy, Morimoto, Wu, Nomura, Yang,
  Shimotani, Suzuki, Toh, Kloc, Xu, Arita, Nagaosa, and Iwasa]{Yuan2013}
Yuan,~H.; Bahramy,~M.~S.; Morimoto,~K.; Wu,~S.; Nomura,~K.; Yang,~B.-J.;
  Shimotani,~H.; Suzuki,~R.; Toh,~M.; Kloc,~C.; Xu,~X.; Arita,~R.; Nagaosa,~N.;
  Iwasa,~Y. Zeeman-type spin splitting controlled by an electric field.
  \emph{Nature Physics.} \textbf{2013}, \emph{9}, 563--569\relax
\mciteBstWouldAddEndPuncttrue
\mciteSetBstMidEndSepPunct{\mcitedefaultmidpunct}
{\mcitedefaultendpunct}{\mcitedefaultseppunct}\relax
\EndOfBibitem
\bibitem[{MacNeill} \latin{et~al.}(2017){MacNeill}, {Stiehl}, {Guimaraes},
  {Buhrman}, {Park}, and {Ralph}]{MacNeill2016}
{MacNeill},~D.; {Stiehl},~G.~M.; {Guimaraes},~M.~H.~D.; {Buhrman},~R.~A.;
  {Park},~J.; {Ralph},~D.~C. {Control of spin-orbit torques through crystal
  symmetry in WTe$_{2}$/ferromagnet bilayers}. \emph{Nature Physics}
  \textbf{2017}, \emph{13}, 300--305\relax
\mciteBstWouldAddEndPuncttrue
\mciteSetBstMidEndSepPunct{\mcitedefaultmidpunct}
{\mcitedefaultendpunct}{\mcitedefaultseppunct}\relax
\EndOfBibitem
\bibitem[Clark \latin{et~al.}(2018)Clark, Neat, Okawa, Bawden,
  Markovi\ifmmode~\acute{c}\else \'{c}\fi{}, Mazzola, Feng, Sunko, Riley,
  Meevasana, Fujii, Vobornik, Kim, Hoesch, Sasagawa, Wahl, Bahramy, and
  King]{Clark2018}
Clark,~O.~J. \latin{et~al.}  Fermiology and Superconductivity of Topological
  Surface States in ${\mathrm{PdTe}}_{2}$. \emph{Phys. Rev. Lett.}
  \textbf{2018}, \emph{120}, 156401\relax
\mciteBstWouldAddEndPuncttrue
\mciteSetBstMidEndSepPunct{\mcitedefaultmidpunct}
{\mcitedefaultendpunct}{\mcitedefaultseppunct}\relax
\EndOfBibitem
\bibitem[Xu \latin{et~al.}(2020)Xu, Wei, Zhou, Feng, Xu, Du, He, Huang, Zhang,
  Liu, Wu, Guo, Wang, Guang, Wei, Peng, Jiang, Yu, and Han]{Xu2020}
Xu,~H. \latin{et~al.}  High Spin Hall Conductivity in Large-Area Type-II Dirac
  Semimetal PtTe2. \emph{Advanced Materials} \textbf{2020}, \emph{32},
  2000513\relax
\mciteBstWouldAddEndPuncttrue
\mciteSetBstMidEndSepPunct{\mcitedefaultmidpunct}
{\mcitedefaultendpunct}{\mcitedefaultseppunct}\relax
\EndOfBibitem
\bibitem[Brown and Beerntsen(1965)Brown, and Beerntsen]{Brown:1965}
Brown,~B.~E.; Beerntsen,~D.~J. {Layer structure polytypism among niobium and
  tantalum selenides}. \emph{Acta Crystallographica} \textbf{1965}, \emph{18},
  31--36\relax
\mciteBstWouldAddEndPuncttrue
\mciteSetBstMidEndSepPunct{\mcitedefaultmidpunct}
{\mcitedefaultendpunct}{\mcitedefaultseppunct}\relax
\EndOfBibitem
\bibitem[Yan \latin{et~al.}(2015)Yan, Cruz, Cook, and Varga]{Yan2015}
Yan,~J.-A.; Cruz,~M. A.~D.; Cook,~B.; Varga,~K. Structural, electronic and
  vibrational properties of few-layer 2H- and 1T-TaSe2. \emph{Scientific
  Reports} \textbf{2015}, \emph{5}\relax
\mciteBstWouldAddEndPuncttrue
\mciteSetBstMidEndSepPunct{\mcitedefaultmidpunct}
{\mcitedefaultendpunct}{\mcitedefaultseppunct}\relax
\EndOfBibitem
\bibitem[M.I.~Aroyo(2016)]{Tables}
M.I.~Aroyo,~E. \emph{International Tables for Crystallography, Volume A,
  Space-group symmetry}; Wiley: New York, 2016\relax
\mciteBstWouldAddEndPuncttrue
\mciteSetBstMidEndSepPunct{\mcitedefaultmidpunct}
{\mcitedefaultendpunct}{\mcitedefaultseppunct}\relax
\EndOfBibitem
\bibitem[Kuhs(1992)]{Kuhs1992}
Kuhs,~W.~F. {Generalized atomic displacements in crystallographic structure
  analysis}. \emph{Acta Cryst. A} \textbf{1992}, \emph{48}, 80--98\relax
\mciteBstWouldAddEndPuncttrue
\mciteSetBstMidEndSepPunct{\mcitedefaultmidpunct}
{\mcitedefaultendpunct}{\mcitedefaultseppunct}\relax
\EndOfBibitem
\bibitem[RuI()]{RuI}
$R_{\text{U}}$ = $\sum \mid \mid I_{\text{obs}}\mid-\mid I_{\text{calc}} \mid
  \mid / \sum \mid I_{\text{obs}}\mid $. Here, $I_{\text{obs}}$,
  $I_{\text{calc}}$ are the experimental and calculated intensities,
  respectively. The summation runs over all relfections. The GOF is given by:
  GOF= $\sqrt{1/(N-P)\cdot\sum [(I_{obs}- I_{calc})^{2}/ \sigma^{2}]}$, where
  the difference between observed and calculated intensities is normalized to
  the uncertainties expressed by the standard deviation ($\sigma$) and to
  $(N-P)$, i.e. the difference between the number of \emph{independent}
  reflections (N) and the number of parameters (P) which are varied.\relax
\mciteBstWouldAddEndPunctfalse
\mciteSetBstMidEndSepPunct{\mcitedefaultmidpunct}
{}{\mcitedefaultseppunct}\relax
\EndOfBibitem
\bibitem[Tusche \latin{et~al.}(2015)Tusche, Krasyuk, and Kirschner]{tus15}
Tusche,~C.; Krasyuk,~A.; Kirschner,~J. Spin resolved bandstructure imaging with
  a high resolution momentum microscope. \emph{Ultramicroscopy} \textbf{2015},
  \emph{159}, 520--529\relax
\mciteBstWouldAddEndPuncttrue
\mciteSetBstMidEndSepPunct{\mcitedefaultmidpunct}
{\mcitedefaultendpunct}{\mcitedefaultseppunct}\relax
\EndOfBibitem
\bibitem[Schneider \latin{et~al.}(2012)Schneider, Wiemann, Patt, Feyer,
  Plucinski, Krug, Escher, Weber, Merkel, Renault, and
  Barrett]{SCHNEIDER2012330}
Schneider,~C.; Wiemann,~C.; Patt,~M.; Feyer,~V.; Plucinski,~L.; Krug,~I.;
  Escher,~M.; Weber,~N.; Merkel,~M.; Renault,~O.; Barrett,~N. Expanding the
  view into complex material systems: From micro-ARPES to nanoscale HAXPES.
  \emph{Journal of Electron Spectroscopy and Related Phenomena} \textbf{2012},
  \emph{185}, 330 -- 339\relax
\mciteBstWouldAddEndPuncttrue
\mciteSetBstMidEndSepPunct{\mcitedefaultmidpunct}
{\mcitedefaultendpunct}{\mcitedefaultseppunct}\relax
\EndOfBibitem
\bibitem[Suga and Tusche(2015)Suga, and Tusche]{SUGA2015119}
Suga,~S.; Tusche,~C. Photoelectron spectroscopy in a wide h$\nu$ region from
  6eV to 8keV with full momentum and spin resolution. \emph{Journal of Electron
  Spectroscopy and Related Phenomena} \textbf{2015}, \emph{200}, 119 --
  142\relax
\mciteBstWouldAddEndPuncttrue
\mciteSetBstMidEndSepPunct{\mcitedefaultmidpunct}
{\mcitedefaultendpunct}{\mcitedefaultseppunct}\relax
\EndOfBibitem
\bibitem[Ryu \latin{et~al.}(2018)Ryu, Chen, Kim, Tsai, Tang, Jiang, Liou, Kahn,
  Jia, Omrani, Shim, Hussain, Shen, Kim, Min, Hwang, Crommie, and Mo]{Ryu2018}
Ryu,~H. \latin{et~al.}  Persistent Charge-Density-Wave Order in Single-Layer
  TaSe2. \emph{Nano Letters} \textbf{2018}, \emph{18}, 689--694, PMID:
  29300484\relax
\mciteBstWouldAddEndPuncttrue
\mciteSetBstMidEndSepPunct{\mcitedefaultmidpunct}
{\mcitedefaultendpunct}{\mcitedefaultseppunct}\relax
\EndOfBibitem
\bibitem[Ge and Liu(2012)Ge, and Liu]{Ge2012}
Ge,~Y.; Liu,~A.~Y. Effect of dimensionality and spin-orbit coupling on
  charge-density-wave transition in 2H-TaSe${}_{2}$. \emph{Phys. Rev. B}
  \textbf{2012}, \emph{86}, 104101\relax
\mciteBstWouldAddEndPuncttrue
\mciteSetBstMidEndSepPunct{\mcitedefaultmidpunct}
{\mcitedefaultendpunct}{\mcitedefaultseppunct}\relax
\EndOfBibitem
\bibitem[Bawden \latin{et~al.}(2016)Bawden, Cooil, Mazzola, Riley,
  Collins-McIntyre, Sunko, Hunvik, Leandersson, Polley, Balasubramanian, Kim,
  Hoesch, Wells, Balakrishnan, Bahramy, and King]{Bawden2016}
Bawden,~L. \latin{et~al.}  Spin-valley locking in the normal state of a
  transition-metal dichalcogenide superconductor. \emph{Nature Communications}
  \textbf{2016}, \emph{7}, 11711\relax
\mciteBstWouldAddEndPuncttrue
\mciteSetBstMidEndSepPunct{\mcitedefaultmidpunct}
{\mcitedefaultendpunct}{\mcitedefaultseppunct}\relax
\EndOfBibitem
\bibitem[Zhou \latin{et~al.}(2016)Zhou, Yuan, Jiang, and Law]{Zhou2016}
Zhou,~B.~T.; Yuan,~N. F.~Q.; Jiang,~H.-L.; Law,~K.~T. Ising superconductivity
  and Majorana fermions in transition-metal dichalcogenides. \emph{Phys. Rev.
  B} \textbf{2016}, \emph{93}, 180501\relax
\mciteBstWouldAddEndPuncttrue
\mciteSetBstMidEndSepPunct{\mcitedefaultmidpunct}
{\mcitedefaultendpunct}{\mcitedefaultseppunct}\relax
\EndOfBibitem
\bibitem[He \latin{et~al.}(2018)He, Zhou, He, Yuan, Zhang, and Law]{He2018}
He,~W.-Y.; Zhou,~B.~T.; He,~J.~J.; Yuan,~N. F.~Q.; Zhang,~T.; Law,~K.~T.
  Magnetic field driven nodal topological superconductivity in monolayer
  transition metal dichalcogenides. \emph{Communications Physics}
  \textbf{2018}, \emph{1}, 104101\relax
\mciteBstWouldAddEndPuncttrue
\mciteSetBstMidEndSepPunct{\mcitedefaultmidpunct}
{\mcitedefaultendpunct}{\mcitedefaultseppunct}\relax
\EndOfBibitem
\bibitem[Sohn \latin{et~al.}(2018)Sohn, Xi, He, Jiang, Wang, Kang, Park,
  Berger, Forro, Law, Shan, and Mak]{Sohn2018}
Sohn,~E.; Xi,~X.; He,~W.-Y.; Jiang,~S.; Wang,~Z.; Kang,~K.; Park,~J.-H.;
  Berger,~H.; Forro,~L.; Law,~K.~T.; Shan,~J.; Mak,~K.~F. An unusual continuous
  paramagnetic-limited superconducting phase transition in 2D NbSe2.
  \emph{Nature Materials} \textbf{2018}, \emph{17}, 504\relax
\mciteBstWouldAddEndPuncttrue
\mciteSetBstMidEndSepPunct{\mcitedefaultmidpunct}
{\mcitedefaultendpunct}{\mcitedefaultseppunct}\relax
\EndOfBibitem
\bibitem[Lu \latin{et~al.}(2015)Lu, Zheliuk, Leermakers, Yuan, Zeitler, Law,
  and Ye]{Lu2015}
Lu,~J.~M.; Zheliuk,~O.; Leermakers,~I.; Yuan,~N. F.~Q.; Zeitler,~U.;
  Law,~K.~T.; Ye,~J.~T. Evidence for two-dimensional Ising superconductivity in
  gated MoS2. \emph{Science} \textbf{2015}, \emph{350}, 1353--1357\relax
\mciteBstWouldAddEndPuncttrue
\mciteSetBstMidEndSepPunct{\mcitedefaultmidpunct}
{\mcitedefaultendpunct}{\mcitedefaultseppunct}\relax
\EndOfBibitem
\bibitem[Saito \latin{et~al.}(2015)Saito, Nakamura, Bahramy, Kohama, Ye,
  Kasahara, Nakagawa, Onga, Tokunaga, Nojima, Yanase, and Iwasa]{Saito2015}
Saito,~Y.; Nakamura,~Y.; Bahramy,~M.~S.; Kohama,~Y.; Ye,~J.; Kasahara,~Y.;
  Nakagawa,~Y.; Onga,~M.; Tokunaga,~M.; Nojima,~T.; Yanase,~Y.; Iwasa,~Y.
  Superconductivity protected by spin-valley locking in ion-gated MoS2.
  \emph{Nature Physics} \textbf{2015}, \emph{12}, 144--149\relax
\mciteBstWouldAddEndPuncttrue
\mciteSetBstMidEndSepPunct{\mcitedefaultmidpunct}
{\mcitedefaultendpunct}{\mcitedefaultseppunct}\relax
\EndOfBibitem
\bibitem[Koepernik and Eschrig(1999)Koepernik, and Eschrig]{Koepernik1999}
Koepernik,~K.; Eschrig,~H. Full-potential nonorthogonal local-orbital
  minimum-basis band-structure scheme. \emph{Phys. Rev. B} \textbf{1999},
  \emph{59}, 1743--1757\relax
\mciteBstWouldAddEndPuncttrue
\mciteSetBstMidEndSepPunct{\mcitedefaultmidpunct}
{\mcitedefaultendpunct}{\mcitedefaultseppunct}\relax
\EndOfBibitem
\bibitem[Eschrig \latin{et~al.}(2004)Eschrig, Richter, and Opahle]{Eschrig2004}
Eschrig,~H.; Richter,~M.; Opahle,~I. In \emph{Relativistic Solid State
  Calculations, in: Relativistic Electronic Structure Theory, Part 2.
  Applications}; Schwerdtfeger,~P., Ed.; Elsevier, 2004; Vol.~13; pp
  723--776\relax
\mciteBstWouldAddEndPuncttrue
\mciteSetBstMidEndSepPunct{\mcitedefaultmidpunct}
{\mcitedefaultendpunct}{\mcitedefaultseppunct}\relax
\EndOfBibitem
\bibitem[Soumyanarayanan \latin{et~al.}(2016)Soumyanarayanan, Reyren, Fert, and
  Panagopoulos]{Soumyanarayanan2016}
Soumyanarayanan,~A.; Reyren,~N.; Fert,~A.; Panagopoulos,~C. Emergent phenomena
  induced by spin-orbit coupling at surfaces and interfaces. \emph{Nature}
  \textbf{2016}, \emph{539}, 509\relax
\mciteBstWouldAddEndPuncttrue
\mciteSetBstMidEndSepPunct{\mcitedefaultmidpunct}
{\mcitedefaultendpunct}{\mcitedefaultseppunct}\relax
\EndOfBibitem
\bibitem[Roy \latin{et~al.}(2014)Roy, Meyerheim, Mohseni, Ernst, Otrokov,
  Vergniory, Mussler, Kampmeier, Gr\"utzmacher, Tusche, Schneider, Chulkov, and
  Kirschner]{Roy.prb2014}
Roy,~S.; Meyerheim,~H.~L.; Mohseni,~K.; Ernst,~A.; Otrokov,~M.~M.;
  Vergniory,~M.~G.; Mussler,~G.; Kampmeier,~J.; Gr\"utzmacher,~D.; Tusche,~C.;
  Schneider,~J.; Chulkov,~E.~V.; Kirschner,~J. Atomic relaxations at the (0001)
  surface of Bi$_2$Se$_3$ single crystals and ultrathin films. \emph{Phys. Rev.
  B} \textbf{2014}, \emph{90}, 155456\relax
\mciteBstWouldAddEndPuncttrue
\mciteSetBstMidEndSepPunct{\mcitedefaultmidpunct}
{\mcitedefaultendpunct}{\mcitedefaultseppunct}\relax
\EndOfBibitem
\bibitem[Cavallin \latin{et~al.}(2016)Cavallin, Sevriuk, Fischer, Manna, Ouazi,
  Ellguth, Tusche, Meyerheim, Sander, and Kirschner]{Cavallin:2016}
Cavallin,~A.; Sevriuk,~V.; Fischer,~K.~N.; Manna,~S.; Ouazi,~S.; Ellguth,~M.;
  Tusche,~C.; Meyerheim,~H.~L.; Sander,~D.; Kirschner,~J. Preparation and
  characterization of Bi2Se3(0001) and of epitaxial FeSe nanocrystals on
  Bi2Se3(0001). \emph{Surface Science} \textbf{2016}, \emph{646}, 72--82\relax
\mciteBstWouldAddEndPuncttrue
\mciteSetBstMidEndSepPunct{\mcitedefaultmidpunct}
{\mcitedefaultendpunct}{\mcitedefaultseppunct}\relax
\EndOfBibitem
\bibitem[Schamper \latin{et~al.}(1993)Schamper, Meyerheim, and
  Moritz]{Schamper1993}
Schamper,~C.; Meyerheim,~H.~L.; Moritz,~W. {Resolution correction for surface
  X-ray diffraction at high beam exit angles}. \emph{Journal of Applied
  Crystallography} \textbf{1993}, \emph{26}, 687--696\relax
\mciteBstWouldAddEndPuncttrue
\mciteSetBstMidEndSepPunct{\mcitedefaultmidpunct}
{\mcitedefaultendpunct}{\mcitedefaultseppunct}\relax
\EndOfBibitem
\bibitem[Vlieg(1997)]{Vlieg1997}
Vlieg,~E. {Integrated Intensities Using a Six-Circle Surface X-ray
  Diffractometer}. \emph{J. Appl. Crystallogr.} \textbf{1997}, \emph{30},
  532--543\relax
\mciteBstWouldAddEndPuncttrue
\mciteSetBstMidEndSepPunct{\mcitedefaultmidpunct}
{\mcitedefaultendpunct}{\mcitedefaultseppunct}\relax
\EndOfBibitem
\bibitem[Tusche \latin{et~al.}(2015)Tusche, Krasyuk, and Kirschner]{Tusche2015}
Tusche,~C.; Krasyuk,~A.; Kirschner,~J. Spin resolved bandstructure imaging with
  a high resolution momentum microscope. \emph{Ultramicroscopy} \textbf{2015},
  \emph{159}, 520--529\relax
\mciteBstWouldAddEndPuncttrue
\mciteSetBstMidEndSepPunct{\mcitedefaultmidpunct}
{\mcitedefaultendpunct}{\mcitedefaultseppunct}\relax
\EndOfBibitem
\bibitem[Tusche \latin{et~al.}(2011)Tusche, Ellguth, Ünal, Chiang, Winkelmann,
  Krasyuk, Hahn, Sch\"{ö}nhense, and Kirschner]{Tusche2011}
Tusche,~C.; Ellguth,~M.; Ünal,~A.~A.; Chiang,~C.-T.; Winkelmann,~A.;
  Krasyuk,~A.; Hahn,~M.; Sch\"{ö}nhense,~G.; Kirschner,~J. Spin resolved
  photoelectron microscopy using a two-dimensional spin-polarizing electron
  mirror. \emph{Applied Physics Letters} \textbf{2011}, \emph{99}, 032505\relax
\mciteBstWouldAddEndPuncttrue
\mciteSetBstMidEndSepPunct{\mcitedefaultmidpunct}
{\mcitedefaultendpunct}{\mcitedefaultseppunct}\relax
\EndOfBibitem
\bibitem[Tusche \latin{et~al.}(2013)Tusche, Ellguth, Krasyuk, Winkelmann,
  Kutnyakhov, Lushchyk, Medjanik, Sch\"onhense, and Kirschner]{Tusche2013}
Tusche,~C.; Ellguth,~M.; Krasyuk,~A.; Winkelmann,~A.; Kutnyakhov,~D.;
  Lushchyk,~P.; Medjanik,~K.; Sch\"onhense,~G.; Kirschner,~J. Quantitative spin
  polarization analysis in photoelectron emission microscopy with an imaging
  spin filter. \emph{Ultramicroscopy} \textbf{2013}, \emph{130}, 70--76\relax
\mciteBstWouldAddEndPuncttrue
\mciteSetBstMidEndSepPunct{\mcitedefaultmidpunct}
{\mcitedefaultendpunct}{\mcitedefaultseppunct}\relax
\EndOfBibitem
\bibitem[Perdew \latin{et~al.}(1996)Perdew, Burke, and Ernzerhof]{Perdew1996}
Perdew,~J.~P.; Burke,~K.; Ernzerhof,~M. Generalized Gradient Approximation Made
  Simple. \emph{Phys. Rev. Lett.} \textbf{1996}, \emph{77}, 3865--3868\relax
\mciteBstWouldAddEndPuncttrue
\mciteSetBstMidEndSepPunct{\mcitedefaultmidpunct}
{\mcitedefaultendpunct}{\mcitedefaultseppunct}\relax
\EndOfBibitem
\end{mcitethebibliography}
\newpage

\begin{figure}[h]
  \center{\includegraphics[width=1.00\columnwidth]{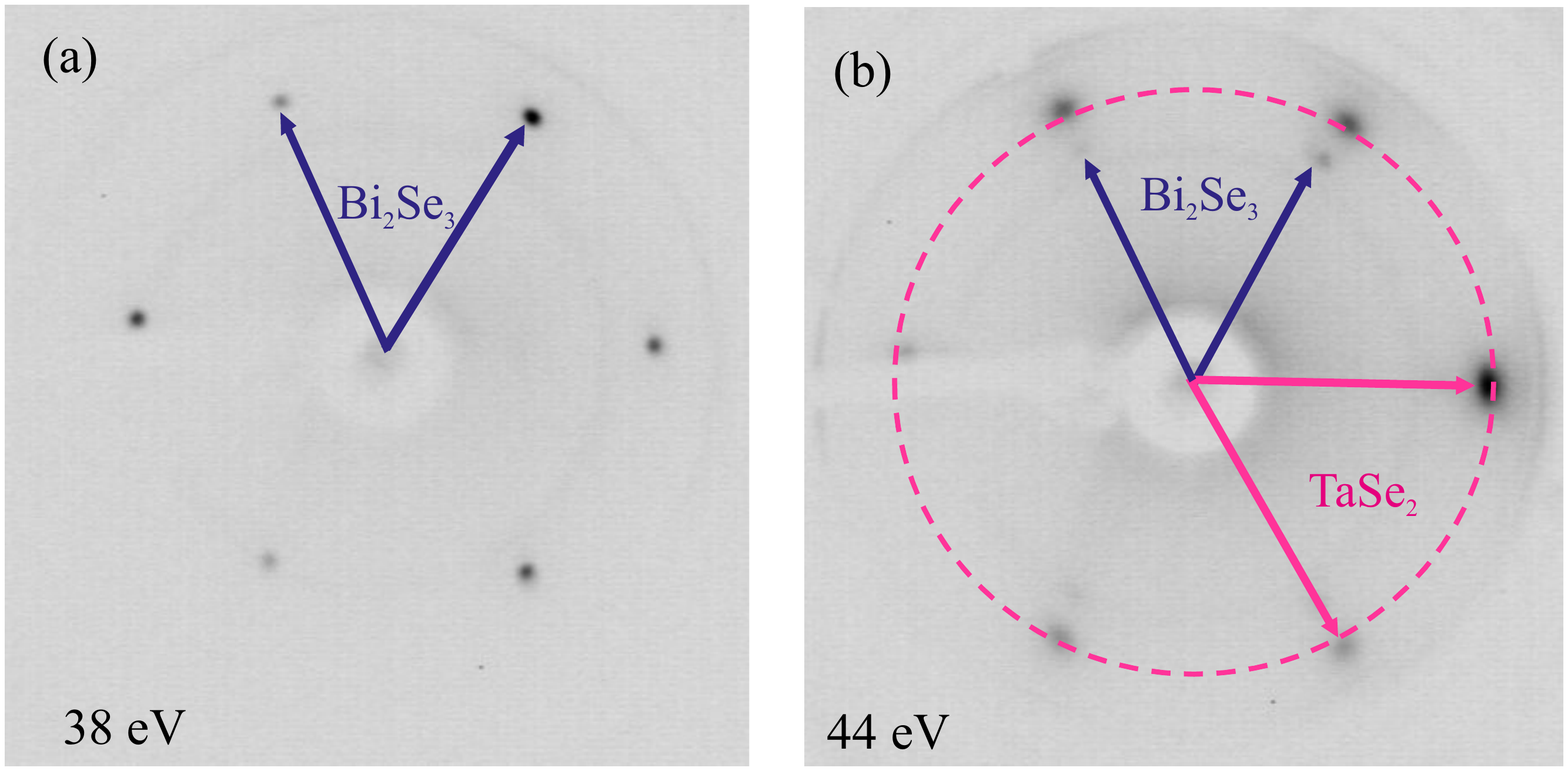}}
  \vspace{0in} \caption{LEED patterns for the pristine (left) and TaSe$_{2}$ covered Bi$_{2}$Se$_{3}$(0001) surface. Note the different symmetries of the LEED patterns: 3m for Bi$_{2}$Se$_{3}$ and p6mm for TaSe$_{2}$, the latter related to the presence of two mutually 60 degrees rotated domains.}
  \label{LEED}
\end{figure}

\begin{figure}[h]
  \center{\includegraphics[width=1.00\textwidth]{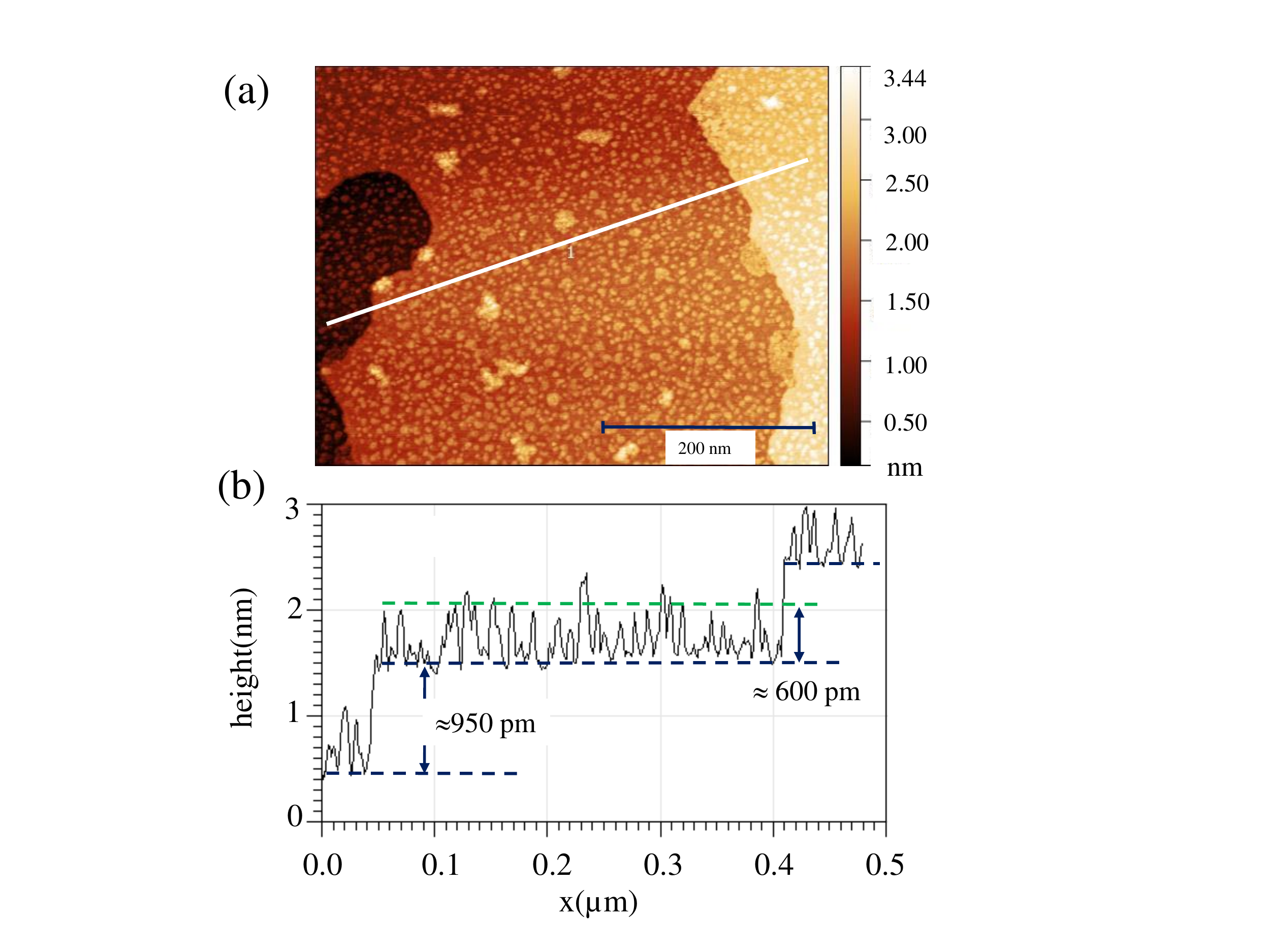}}
  \vspace{0in} \caption{(a): Scanning tunnelling microscopy (U=-1V, I=100 pA) image of TaSe$_{2}$ islands (bright) on the Bi$_{2}$Se$_{3}$(0001) surface. (b): Height profile along the white line in (a). Note, the step height of the terrace ($\approx$950~pm) related to a full QL. The height of the islands is approximately equal to 600~pm related to the height of a single Se-Ta-Se monosheet on the Bi$_{2}$Se$_{3}$ surface.}
  \label{STM}
\end{figure}

\begin{figure}[h]
  \center{\includegraphics[width=1.00\textwidth]{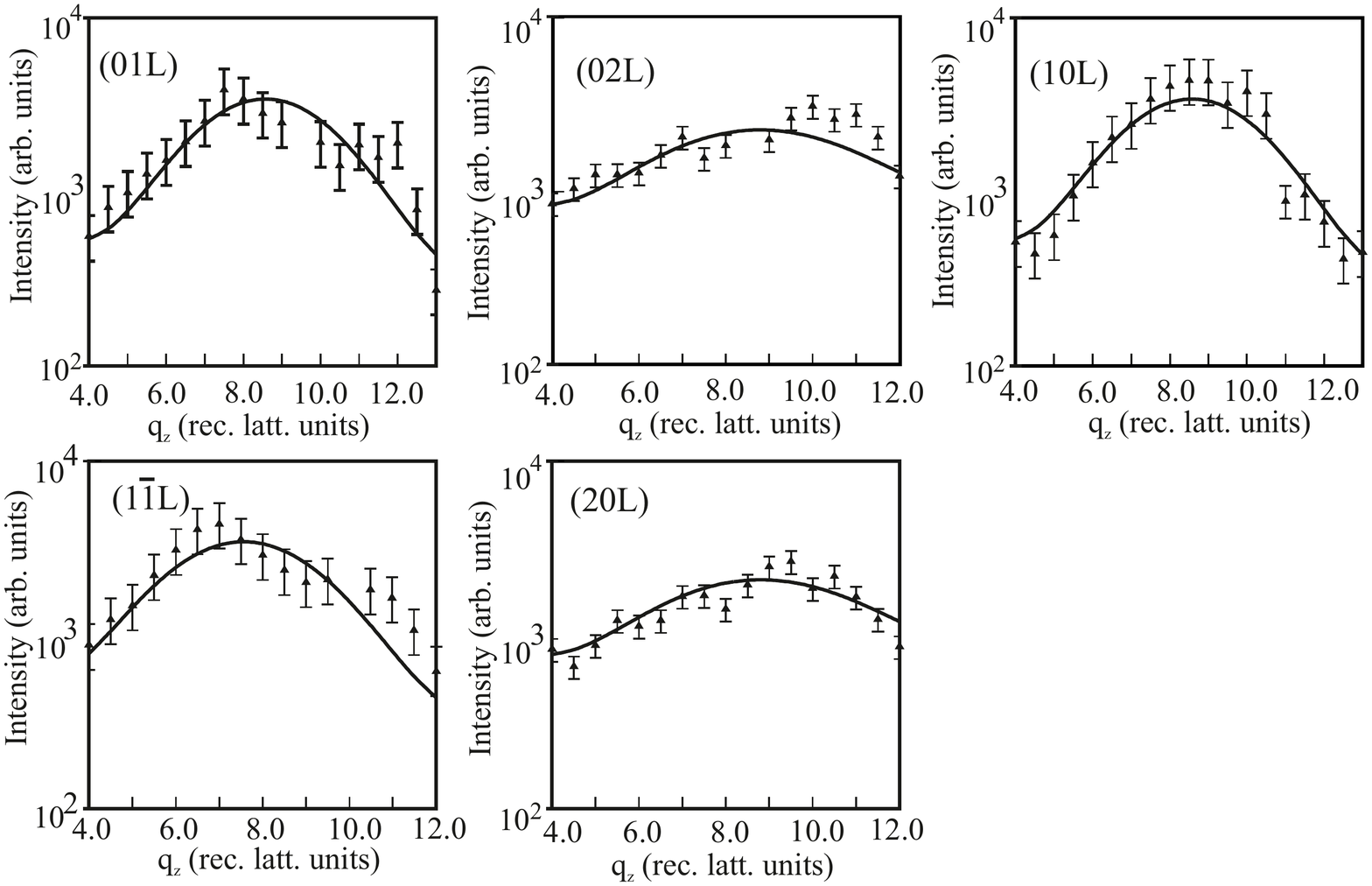}}
  \vspace{0in} \caption{Experimental (symbols) and calculated (lines) intensities for a monosheet of TaSe$_{2}$ on Bi$_{2}$Se$_{3}$(0001) along several symmetry independent rods in reciprocal lattice as labelled. The unit of the perpendicular momentum transfer (q$_{z}$) is referred to the Bi$_{2}$Se$_{3}$ substrate lattice (1/c$_{0}$=0.349nm$^{-1}$.)}
  \label{Rods}
\end{figure}

\begin{figure}[h]
  \center{\includegraphics[width=1.00\textwidth]{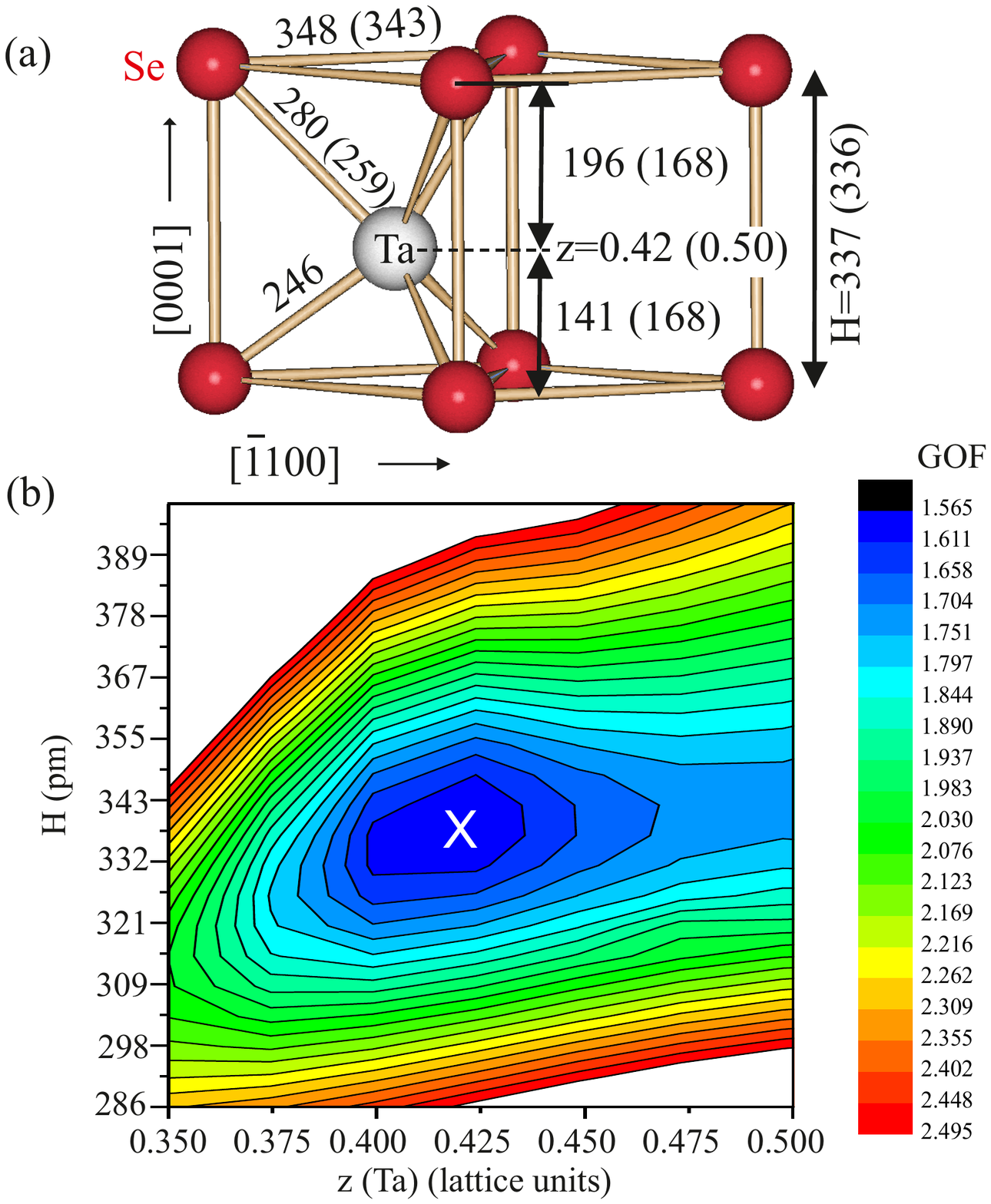}}
  \vspace{0in} \caption{(a): Structure model of the H-TaSe$_{2}$ monosheet derived from the SXRD analysis. Red and grey spheres represent selenium and tantalum, respectively. Numbers indicate distances in picometer (pm) units with those in brackets referring to the bulk crystal. (b): Contour plot of GOF versus H (height of the prismatic monosheet) and z (relative tantalum position within the prismatic unit. The global minimum is given by the cross at H=337 pm and z=0.42. One contour level corresponds to a step in  GOF by 3\%. Uncertainties of H and z are estimated on the basis of $\Delta$ GOF=+3\%. For details see text.}
  \label{Structure}
\end{figure}

\begin{figure}[h]
  \center{\includegraphics[width=0.950\textwidth]{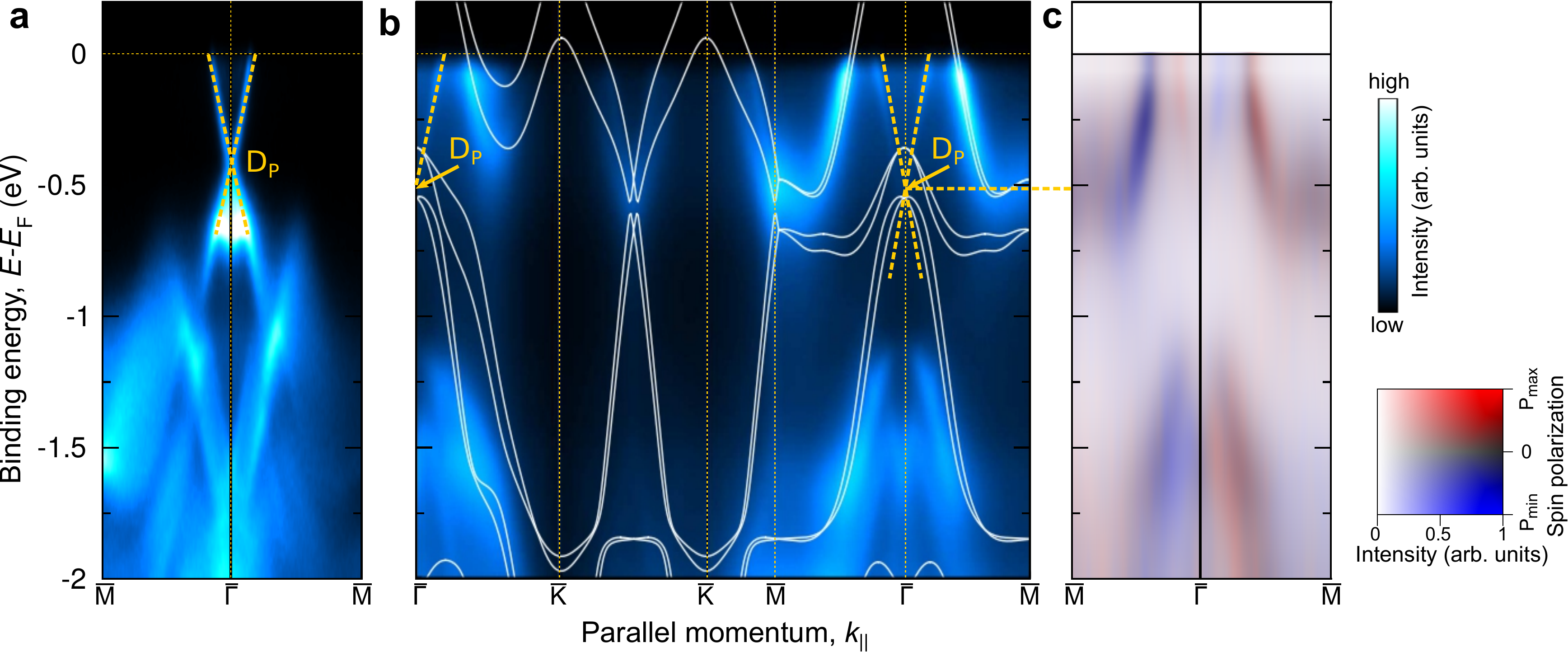}}
  \vspace{0in} \caption{Experimental spin-averaged band structure of the pristine (a) and the TaSe$_{2}$ covered (b) Bi$_{2}$Se$_{3}$(0001) surface along directions connecting high-symmetry points in the first BZ. Lines in (b) represent the calculation based on the SXRD structure model. The spin-resolved band structure of the TaSe$_{2}$ covered surface is shown in (c). The spectral density and spin-polarization projected along the y-axis are represented by the scale bars on the right.}
  \label{ARPES1}
\end{figure}

\begin{figure}[h]
  \center{\includegraphics[width=0.95\textwidth]{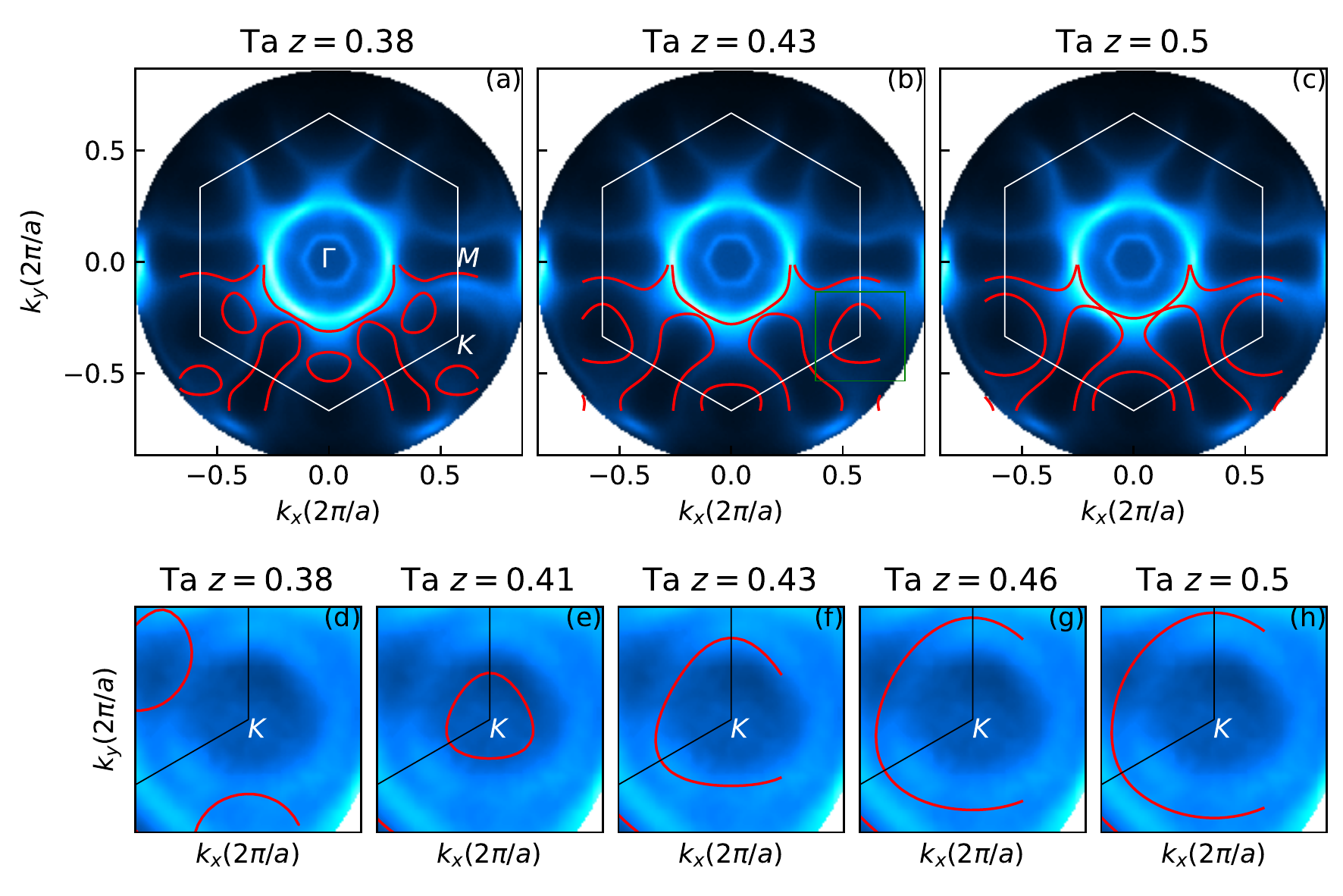}}
  \vspace{0in} \caption{(a)-(c): Comparison between the experimental photoemission momentum map of TaSe$_{2}$/Bi$_{2}$Se$_{3}$ at the Fermi Surface with calculated ones (red lines) for positions z=0.38, 0.43 and 0.50 of the tantalum atom within the Se-Ta-Se triple layer. The white hexagon indicates the first Brilluoin zone. (c)-(h): close up of the hole pocket at the $\overline{K}$-point emphasizing the dependence of its size and position on z. The best match is observed for z being in the range 0.41 to 0.43 in agreement with SXRD.}
  \label{ARPESCALC}
\end{figure}

\begin{figure}[h]
  \center{\includegraphics[width=0.90\textwidth]{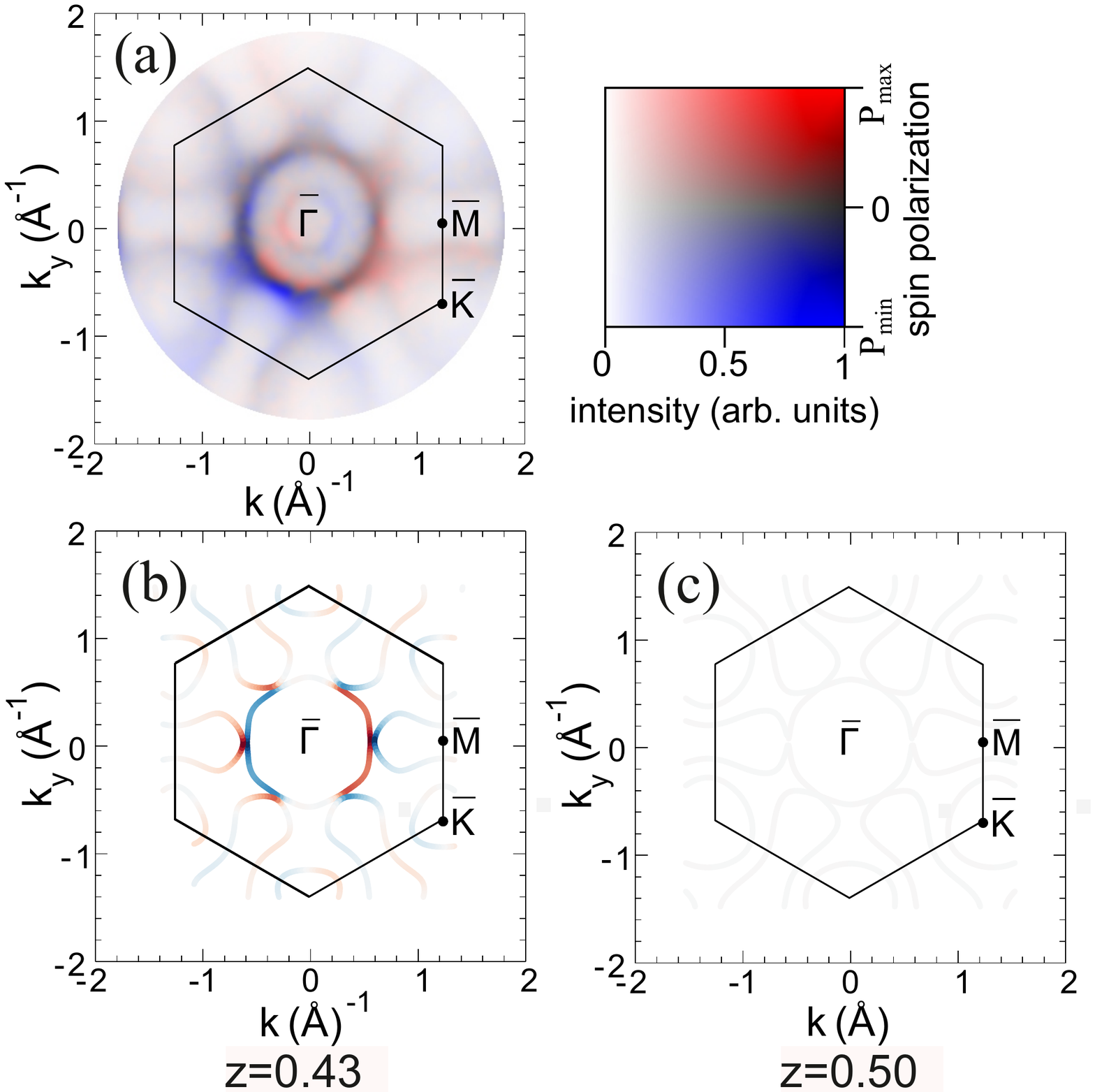}}
  \vspace{0in} \caption{ Experimental spin-resolved photoemission momentum map of TaSe$_{2}$ monosheet on Bi$_{2}$Se$_{3}$ at the Fermi level (a) compared with calculations in which the position of the tantalum atom is located at z=0.43 (b), z=0.50 (c). Color code quantifies the degree of the in-plane polarization along the y-axis as given by the scale bar on the right.}
  \label{ARPESIPL}
\end{figure}


\end{document}